\documentclass[a4paper,pra,showpacs,superscriptaddress,groupedaddress]{revtex4}
\usepackage{ae}
\usepackage{physics}
\usepackage[T1]{fontenc}
\usepackage[ansinew]{inputenc}
\usepackage{amsmath}
\usepackage{amssymb}
\usepackage[caption=false]{subfig}
\usepackage{float}
\usepackage{multirow}
\usepackage{array}
\usepackage[]{graphicx}
\usepackage{wrapfig}
\usepackage{makecell}
\usepackage{mathtools}
\usepackage{color}
\usepackage[colorlinks]{hyperref}

\usepackage{lscape}
\newtheorem{theorem}{Theorem}
\graphicspath{ {./image/} }

\begin{document}
\title{Multipartite Monogamy of Entanglement for Three Qubit States}
\author{Priyabrata Char}
\email{mathpriyabrata@gmail.com}
\affiliation{Department of Applied Mathematics, University of Calcutta, 92 A.P.C Road, Kolkata, 700009, West Bengal, India}
\author{Dipayan Chakraborty}
\email{dipayan.tamluk@gmail.com}
\affiliation{Department of Mathematics, Sukumar Sengupta Mahavidyalaya, Keshpur, Paschim Medinipur, 721150, West Bengal, India}
\author{Prabir Kumar Dey}
\email{prabirkumardey1794@gmail.com}
\affiliation{Department of Science and Humanities, Sree Ramkrishna Silpa Vidyapith, Suri, Birbhum, 731101, West Bengal, India}
\author{Ajoy Sen}
\email{ajoy.sn@gmail.com}
\affiliation{Department of  Mathematics, East Calcutta Girls' College, P-237 Lake Town Link Road, Kolkata 700089, West Bengal, India}
\author{Amit Bhar}
\email{bharamit79@gmail.com}
\affiliation{Department of Mathematics, Jogesh Chandra Chaudhuri College, 30, Prince Anwar Shah Road, Kolkata, 700033, India}
\author{Indrani Chattopadhyay}
\email{icappmath@caluniv.ac.in}
\affiliation{Department of Applied Mathematics, University of Calcutta, 92 A.P.C Road, Kolkata, 700009, West Bengal, India}
\author{Debasis Sarkar}
\email{dsarkar1x@gmail.com, dsappmath@caluniv.ac.in}
\affiliation{Department of Applied Mathematics, University of Calcutta, 92 A.P.C Road, Kolkata, 700009, West Bengal, India}

\begin{abstract}
The distribution of entanglement in a multiparty system can be described through the principles of monogamy or polygamy. Monogamy is a fundamental characteristic of entanglement that restricts its distribution among several  number of parties(more than two). In this work, our aim is to explore how quantum entanglement can be distributed in accordance with monogamy relations by utilizing both the genuine multipartite entanglement measures and bipartite entanglement measures. Specifically, we treat source entanglement as the genuine multipartite entanglement measure and use the entanglement of formation specifically for bipartite cases. For GHZ class states, we analytically demonstrate that the square of the source entanglement serves as an upper bound for the sum of the squares of the entanglement of formation of the reduced subsystems, with some exceptions for specific non-generic GHZ states. We also present numerical evidence supporting this result for W class states. Additionally, we explore the monogamy relation by using accessible entanglement as an upper bound.
\end{abstract}
\date{\today}
\maketitle
\section{Introduction}
Composite systems, together with correlations, give rise to several non-trivial and striking phenomena in different areas of quantum theory. The correlations of quantum states help us visualize the physics of many particle systems more profoundly. Entanglement is a special type of quantum correlation, and multipartite entanglement is regarded as a multilinear resource \cite{ LOCC1, LOCC2}. LOCC (Local Operation along with Classical Communication) \cite{LOCC5, LOCC6, LOCC7, Atanu 3, Atanu 5} is considered as the type of quantum operation that consumes and manipulates the entanglement. The non-increasing feature of entanglement under LOCC is considered the thermodynamic law of entanglement and has a great impact on the manipulation of entanglement for the proper implementation of various quantum information tasks \cite{LOCC3, LOCC4}. Although LOCC has a clear physical description, the mathematical characterization \cite{math} of LOCC is very hard. The existence of maximally entangled states under LOCC is to be posed as not only executing the quantum protocols optimistically but also getting more advantages from these protocols \cite{Atanu 1, Atanu 2, Atanu 4}. Different quantum information protocols and Nielsen's majorization condition \cite{5} for deterministic bipartite pure state transformation under LOCC established that the maximally entangled state in a bipartite quantum system is unique. However, the scenario has drastically changed from a three-qubit system. The existence of more than one SLOCC (Stochastic Local Operations with Classical Communications) inequivalent class \cite{6,7,8,9} is one of the main restrictions on the presence of a global maximally entangled state in multipartite quantum systems. Each SLOCC class contains a corresponding maximally entangled state (up to local unitary). To cope with this problem, J.I. de Vicente et al. introduced \cite{10,11} the concept of $\text{MES}_n$ (maximally entangled set of $n$-partite states). The $\text{MES}_n$ is the smallest set of $n$-party states such that any other truly $n$-partite entangled state can be deterministically obtained via LOCC from a state in $\text{MES}_n$.
Most of the known entanglement measures are difficult to compute in a multipartite scenario and lack interpretation in terms of LOCC convertibility. Keeping this in mind, the idea of source entanglement($E_s$) and accessible entanglement($E_a$) has been put forward by Schwaiger et al.\cite{12}. The main idea was to incorporate more physical single-copy entanglement transformation instead of the usual asymptotic limit of many copies. These two measures have been computed for pure three-qubit \cite{12}, four-qubit systems \cite{Kraus 1}, and some low-dimensional bipartite systems \cite{Kraus 1}. Source entanglement$(E_s)$ measures the volume of states from which a given state can be reached via LOCC. On the other hand, accessible entanglement$(E_a)$ measures the volume of states to which we can reach from a given state via LOCC.

Monogamy is an interesting feature of entanglement that prohibits the free sharing of resources among parties, unlike classical correlation.
The monogamy relations of entanglement reveal the distribution pattern of entanglement within a composite quantum system. This concept was first introduced by Coffman, Kundu, and Wooters in their seminal paper \cite{1}. They mathematically formulated monogamy as an inequality involving concurrence \cite{2}.The CKW inequality \cite{1} is given by
\begin{equation}
\label{CKW}
C^{2}_{A|BC}\geq C^{2}_{AB}+C^{2}_{AC}.
\end{equation}
They also introduced a new multipartite entanglement measure known as residual entanglement, or tangle \cite{1, M1}, defined as
\begin{equation}
\tau= C^{2}_{A|BC}- C^{2}_{AB}-C^{2}_{AC}.
\end{equation}
Generalization of inequality \eqref{CKW} has been perfectly done by Osborne and Verstrate \cite{generalCKW1} for $n$ qubit pure states. Similar types of inequality are also satisfied by Negativity \cite{M5}, Squared Entanglement of formation \cite{SEOF monogamy 1, SEOF monogamy 2}, squashed entanglement \cite{squashed monogamy 1, squashed monogamy 2}, one-way distillable entanglement \cite{squashed monogamy 1}, R\'enyi-$\alpha$ entropy \cite{Renyi monogamy 1, Renyi monogamy 2}, squared R\'enyi-$\alpha$ entropy \cite{Squared Renyi monogamy 1}, and Tsallis-$q$ entanglement \cite{Tsallis monogamy 1, Tsallis monogamy 2}. However, there are instances where monogamy is violated, such as with the entanglement of formation \cite{M9}. Various works \cite{M2, M3, M4, M6, M7, M8, M9, M11, M12, M13, char exp monogamy} have been done to find varieties of monogamy relations during the last two decades. Application of this feature has already been traced in different areas of quantum information theory such as frustrated spin systems \cite{generalCKW1, frustrated}, quantum key distribution \cite{keydistribution}, black hole theory \cite{black}, etc. In \cite{3}, Marcio F. Cornelio has derived a monogamy relation for three-qubit pure and mixed states
\begin{equation}
C_{3}^{2}(\rho_{ABC})\geq C^2(\rho_{AB})+C^2(\rho_{AC})+C^2(\rho_{BC}).
\end{equation}
Since the upper bound $C_{3}^{2}(\rho_{ABC})$ represents multiparty entanglement \cite{generalized concurrence 1, generalized concurrence 2}, the inequality discussed is termed as a multiparty monogamy relation \cite{3}. In \cite{13}, the authors have discussed relations between source entanglement and entanglement of formation of bipartite states. In this study, we will examine two multiparty entanglement measures (source entanglement and accessible entanglement) with the entanglement of formation in the context of monogamy relations, focusing specifically on three-qubit pure states. Our primary analysis involves a pure 3-qubit system, using squared entanglement of formation to evaluate bipartite entanglement of reduced systems and assessing multipartite entanglement through source entanglement $(E_s)$ or accessible entanglement $(E_a)$. Thus, our monogamy inequalities are
\begin{equation}
\label{M1}
E_{s}^{2}\geq E_{12}^{2}+E_{23}^{2}+E_{13}^{2},
\end{equation}
\begin{equation}
\label{M2}
E_{a}^{2}\geq E_{12}^{2}+E_{23}^{2}+E_{13}^{2},
\end{equation}

where $E_s$ denotes the source entanglement, $E_a$ denotes the accessible entanglement and $E_{ij}$ is the entanglement of formation of $ij-th$ ($i\neq j$ and $i,j=1,2,3$) subsystems of the state. These are multipartite monogamy relations, as the upper bounds $E_{s}$ and $E_a$ are genuine multipartite entanglement measures. We will consider the monogamy scores $M_1=E_{s}^{2}-E_{12}^{2}-E_{13}^{2}-E_{23}^{2}$ and $M_2=E_{a}^{2}-E_{12}^{2}-E_{13}^{2}-E_{23}^{2}$ while discussing monogamy. Whenever $M_1$ or $M_2$ are non-negative, we can say the corresponding monogamy relation (\ref{M1}) or (\ref{M2}) is satisfied.

Our paper is structured as follows: Section II covers the essential concepts needed for our analysis. In Section III, we examine the monogamy relations for pure GHZ class states in three-qubit systems. Section IV is dedicated to discussing multipartite monogamy relations for pure W class states in three-qubit systems. Section V presents our observations and interpretations of these monogamy relations, and we conclude with a summary in Section VI.

\section{Preliminary Ideas}
\subsection{Concurrece}
For a two-qubit state $\rho_{AB}$, concurrence \cite{2} is an important entanglement monotone and it is defined as 
\begin{equation}
C(\rho_{AB})=\max\{0,\lambda_1-\lambda_2-\lambda_3-\lambda_4\},
\end{equation}
where $\lambda_1, \lambda_2, \lambda_3, \lambda_4$ are the square roots of the eigen values of the matrix $\rho_{AB}((\sigma_y\otimes\sigma_y)\rho^{*}_{AB}(\sigma_y\otimes\sigma_y))$ in decreasing order, where $\sigma_y$ is the Pauli spin matrix and $\rho^{*}_{AB}$ is conjugate of $\rho_{AB}$. For a pure bipartite state, the concurrence can be computed as 
\begin{equation}
C(\rho_{AB})=2\sqrt{\det\rho_A},
\end{equation}
 where $\rho_A$ is obtained from $\rho_{AB}$ by taking partial trace over qubit B. 
 Entanglement of formation \cite{2} is another entanglement measure of $\rho_{AB}$ is defined as
\begin{equation}
\label{Ef}
E(\rho_{AB})=-x\log_2x-(1-x)\log_2(1-x),
\end{equation}
where $x=\frac{1+\sqrt{1-C^{2}(\rho_{AB})}}{2}$. If the concurrence of a reduced state is zero, then its entanglement of formation will also be zero. From here on, we will use the symbols $C_{AB}$ and $E_{AB}$ instead of $C(\rho_{AB})$ and $E(\rho_{AB})$ respectively. The generalization of concurrence is not unique, and we will consider the form as introduced in \cite{generalized concurrence 1, generalized concurrence 2} by Carvalho et al. For an $N$ partite state $\Phi_N$ it is defined as
\begin{equation}
C_N(\Phi_N)=2^{1-\frac{N}{2}}\sqrt{(2^N-2)-\sum_{i}\tr\rho_{i}^{2}},
\end{equation}
where the sum runs over all $2^N-2$ subsystems of the given $N$ partite system. 
 
\subsection{Source and accessible entanglement}
In 2015, K. Schwaiger et al. \cite{12} introduced two novel operational entanglement measures for multipartite states (whether pure or mixed) of arbitrary dimensions. These measures can be computed given that knowledge of all possible LOCC transformations is known, termed source entanglement($E_s$) and accessible entanglement($E_a$). If a state $\ket{\psi}$ can reach a state $\ket{\phi}$ using a LOCC protocol, we say that $\ket{\phi}$ is accessible from $\ket{\psi}$. For a given state $\ket{\psi}$, let $M_a(\ket{\psi})$ represent the set of states that can be accessed from  $\ket{\psi}$ via LOCC. Similarly, $M_s(\ket{\psi})$ denotes the set of states that can reach $\ket{\psi}$ via LOCC. Clearly, $M_s(\ket{\psi})\subseteq M_s(\ket{\phi})$ and $M_a(\ket{\phi})\subseteq M_a(\ket{\psi})$ whenever $\ket{\phi}$ is accessible from $\ket{\psi}$ through LOCC. Thus, the monotonic properties of $M_a$ and $M_s$ under LOCC transformations are established. This indicates that any normalized and appropriately scaled measure of these sets can serve as an entanglement measure. In \cite{12}, the authors defined the accessible volume as
\begin{equation}
 \label{eq:accessiblevolume}
 V_a(\ket{\psi})=\mu(M_a(\ket{\psi})),
 \end{equation} 
 and source volume as 
 \begin{equation}
  \label{eq:sourcevolume}
 V_s(\ket{\psi})=\mu(M_s(\ket{\psi})),
 \end{equation} where $\mu$ is an arbitrary measure in the set of local unitary equivalence classes. Now, accessible entanglement is defined as
\begin{equation}
 \label{eq:accessibleentanglement}
E_a(\ket{\psi})=\frac{V_a(\ket{\psi})}{V_{a}^{\sup}},
\end{equation}
and the source entanglement is defined as 
\begin{equation}
\label{eq:sourceentanglement}
E_s(\ket{\psi})=1-\frac{V_s(\ket{\psi})}{V_{s}^{\sup}},
\end{equation}
where $V_{a}^{\sup}$ and $V_{s}^{\sup}$ denote the maximum accessible and source volume according to the measure $\mu$. Considering the Lebesgue measure and using convex geometry, the source and accessible volume have been obtained for three, four-qubit and low-dimensional bipartite systems.

\section{Multipartite monogamy in GHZ class states}
Before we delve into the direct computation of source entanglement and other measures for GHZ class states, we will first review the physical and mathematical properties of these states. It's crucial to understand that the true nature of tripartite entanglement cannot be fully captured by analyzing the entanglement of its individual subsystems alone. The state $\ket{GHZ} =\frac{1}{\sqrt{2}}(\ket{000}+\ket{111})$ is considered a generalization of the Bell state from two to three-qubit system. Any state in the GHZ SLOCC class can be written as (up to local unitaries (LUs)) $\ket{\psi(\vec{g},z)}=\frac{1}{\sqrt{k}}g_{x}^{1}\otimes g_{x}^{2}\otimes g_{x}^{3}P_z (\ket{000}+\ket{111})$ where
$$g_{x}^{i}=
\begin{pmatrix}
\frac{1}{\sqrt{2}} & \sqrt{2}g_i\\
0 & \frac{1}{\sqrt{2}}\sqrt{1-4g_{i}^{2}}
\end{pmatrix},
$$
so that $(g_{x}^{i})^\dagger(g_{x}^{i})=\frac{1}{2}I+g_i\sigma_x$, $g_i\in[0,\frac{1}{2})$, $\forall i=1(1)3$, $\vec{g}=(g_1,g_2,g_3)$, $P_z=\begin{pmatrix}
z & 0\\
0 & 1/z
\end{pmatrix}$,
$z\in\mathbb{C}$ with $|z|\leq1$ and $\frac{1}{\sqrt{k}}$ is the normalizing factor, $k=\frac{1}{8|z|^2}\{1+|z|^4+(z^2+(z^{*})^{2})8g_1g_2g_3\}$.
Now the concurrences of the reduced states of $\ket{\psi(\vec{g},z)}$ are,
\begin{equation}
\label{c12}
C_{12}=\frac{2g_3\sqrt{1-4g_{1}^{2}}\sqrt{1-4g_{2}^{2}}}{4k}
\end{equation}
\begin{equation}
\label{c13}
C_{13}=\frac{2g_2\sqrt{1-4g_{1}^{2}}\sqrt{1-4g_{3}^{2}}}{4k}
\end{equation}
\begin{equation}
\label{c23}
C_{23}=\frac{2g_1\sqrt{1-4g_{2}^{2}}\sqrt{1-4g_{3}^{2}}}{4k}
\end{equation}
The following theorem \cite{10, 12} classifies the GHZ state within $\text{MES}_3$.
\begin{theorem}
A state from the GHZ class is in $\text{MES}_3$ if and only if (i) $z=\pm1$, (ii) either none of $g_1,\;g_2,\;g_3$ vanishes or all three of them vanish. \cite{10}
\end{theorem}
The three qubit pure states from GHZ class that are not part of $MES_3$ and for which none of the parameters $g_1,\;g_2,\;g_3$ vanish, are called generic GHZ states \cite{12}. If at least one of the parameters $g_i=0$ for $i=1,2,3$, then they are called non-generic GHZ states \cite{12}. Non-generic three-qubit GHZ pure states allow us to treat the parameter $z$ as a real number $r\in(0,1]$ since a local unitary transformation can be applied to the qubit with the vanishing parameter $g_i$ to absorb the phase of $z$.

Although states from the GHZ class that belong to $\text{MES}_3$ and have all three parameters
$g_1, g_2, g_3$ are zero (LU equivalent to GHZ state) are included in non-generic GHZ states, pure GHZ class states in $\text{MES}_3$ with non-zero parameters $g_1, g_2, g_3$ do not fall into either the generic or non-generic categories. These states have the form $\pm\frac{1}{\sqrt{k}}g_{x}^{1}\otimes g_{x}^{2}\otimes g_{x}^{3} \ket{GHZ}$ where $g_i\neq 0 ,\forall i=1,2,3$.

The following three theorems \cite{12} describe the LOCC transformations between states within the GHZ SLOCC class. These results enable us to compute the source and accessible volumes of both non-generic and generic GHZ states.

\begin{theorem}
\label{th:generictran}
    State transformation via LOCC form $\ket{\psi(g_1,g_2,g_3,z)}$ to  $\ket{\psi(h_1,h_2,h_3,z')}$ where $g_i,h_i\neq 0 \forall i=1,2,3$ is possible iff (i) $g_i\leq h_i, \forall i=1,2,3$ (ii) $\frac{g_1g_2g_3}{h_1h_2h_3}=\frac{Re(z'^2)}{1+|z'|^2}\frac{1+|z|^4}{Re(z^2)}=\frac{Im(z'^2)}{|z'|^2-1}\frac{|z|^4-1}{Im(z^2)}$.   
\end{theorem}
\begin{theorem}
\label{th:nongenerictran1}
        State transformation via LOCC form $\ket{\psi(g_1,g_2,g_3,z)}$ to  $\ket{\psi(h_1,h_2,h_3,z')}$ where $g_i$ is arbitrary and at least one $h_i=0$ is possible iff (i) $g_i\leq h_i,\forall i=1,2,3$, (ii) $r\geq r'$ where $r=|z| \text{, } r'=|z'|$.    
\end{theorem}

\begin{theorem}
\label{th:nongenerictran2}
        State transformation via LOCC form $\ket{\psi(g_1,g_2,g_3,z)}$ to  $\ket{\psi(h_1,h_2,h_3,z')}$ where at least one $g_i=0$ and all $h_i\neq 0$ is possible iff (i) $g_i\leq h_i\forall i=1,2,3$, (ii) $r=1$, (iii) $z'=r'e^{i\phi'}$ with $\phi'=\frac{\pi}{4},\frac{3\pi}{4}$ and $r'$ is arbitrary.  
\end{theorem}

Using theorem (\ref{th:nongenerictran1}), it is straightforward to calculate the source and accessible volume of $\ket{\psi(0,0,0,r)}$ 
$$V_s({\psi(0,0,0,r)}=\int_{r}^{1}dr'=(1-r)$$ $$V_a({\psi(0,0,0,r)})=3\int_{0}^{\frac{1}{2}}\int_{0}^{\frac{1}{2}}\int_{0}^{r}dr'dh_2dh_3=\frac{3r}{4}.$$ 
Using similar argument of theorem (\ref{th:nongenerictran1}), source and accessible volume of $\ket{\psi(g_1,0,0,r)}$ and $\ket{\psi(g_1,g_2,0,r)}$ can be calculated as 
\begin{equation*}
\begin{split}
V_s(\ket{\psi(g_1,0,0,r)})=\int_{0}^{g_1}\int_{r}^{1}dr'dh_1=g_1(1-r),\\
V_a(\ket{\psi(g_1,0,0,r)})=2\int_{0}^{\frac{1}{2}}\int_{g_1}^{\frac{1}{2}}\int_{0}^{r}dr'dh_1dh_2=(\frac{1}{2}-g_1)r,\\
\end{split}
\end{equation*}
\begin{equation*}
\begin{split}
V_s(\ket{\psi(g_1,g_2,0,r)})& =\int_{0}^{g_1}\int_{0}^{g_2}\int_{r}^{1}dr'dh_1dh_2\\ &=g_1g_2(1-r),\\
\end{split}
\end{equation*}
\begin{equation*}
\begin{split}
V_a(\ket{\psi(g_1,g_2,0,r)})&=\int_{g_1}^{\frac{1}{2}}\int_{g_2}^{\frac{1}{2}}\int_{0}^{r}dr'dh_1dh_2\\&=(\frac{1}{2}-g_1)(\frac{1}{2}-g_2)r.\\
\end{split}
\end{equation*}
When $r=1$, we can  calculate source and accessible volume for the states $\ket{\psi(0,0,0,1)},\ket{\psi(g_1,0,0,1)}$ and $\ket{\psi(g_1,g_2,0,1)}$ using theorem (\ref{th:nongenerictran2}). For the sate $\ket{\psi(g_1,g_2,g_3,r)}$, theorem  (\ref{th:generictran})  provides the necessary calculations for the source and accessible volumes. Detailed computations for the source and accessible volumes of GHZ SLOCC class states can be found in \cite{12}. We will now discuss our results for three-qubit GHZ class states in detail. Our exploration covers the monogamy relations across two primary subclasses of GHZ class states, as well as other GHZ class states that belong to $MES_3$. These are: A. Non-generic states in GHZ class, B. Generic states in GHZ class, and C. State from $MES_3$ of the form $\frac{1}{\sqrt{k}}g_{x}^{1}\otimes g_{x}^{2}\otimes g_{x}^{3} \ket{GHZ}$ where $g_1,g_2,g_3 \neq 0$.

 \subsection{Non-Generic GHZ states}

In this subsection, we will establish some monogamy relations for the GHZ class of non-generic pure states. These non-generic GHZ class states are of particular interest because the entanglement of the reduced bipartite system depends on the $g_i$ parameter, as shown by the relations \eqref{c12}-\eqref{c23} can be zero. Although the reduced bipartite system may or may not be entangled, each case exhibits a non-zero tangle. Given the direct relationship between $g_i$ and the entanglement of the reduced subsystem, we have thoroughly explored our results by considering the following three possibilities.\\

\textbf{Case 1 ($g_1=0$, $g_2=0$, $g_3=0$) :} In this subclass, we encounter the simplest structure of the non-generic GHZ class, characterized by the absence of bipartite entanglement in all the reduced subsystems. Utilizing the results from equations \eqref{c12}-\eqref{c23} it can be shown that $C_{ij}=0$ and hence $E_{ij}=0$ for $i\neq j$ and $i,j=1,2,3$. The source entanglement and accessible entanglement of the states belonging to this subclass are given by,
 $$
 E_s=E_a= 
 \begin{cases} 
 r &  \textrm{if  }r \in (0,1) \\  
 1 & \textrm{if  }r=1
 \end{cases}
 $$ 
 Hence the monogamy relations $E_{s}^{2}\geq E_{12}^{2}+E_{13}^{2}+E_{23}^{2}$ and 
 $E_{a}^{2}\geq E_{12}^{2}+E_{13}^{2}+E_{23}^{2}$ hold trivially in this case.\\
 
\textbf{Case 2 ($g_1\neq 0$, $g_2=0$, $g_3=0$) :} In this non-generic GHZ class, two parameters, $g_2$ and $g_3$, are considered to be zero.  As a result, the entanglement of the reduced subsystems and the structure of the state undergo significant changes compared to the previous case. The concurrences (refer eqns. \eqref{c12}-\eqref{c23}) are $C_{12}=C_{13}=0 $ $\textrm{and } C_{23}=\frac{2g_1}{4k}$ and hence $E_{12}=E_{13}=0$ and $E_{23}\neq 0$. The source and accessible entanglement in this case are given by\\
 
\begin{figure} [h]
\centering
\includegraphics[scale=0.5]{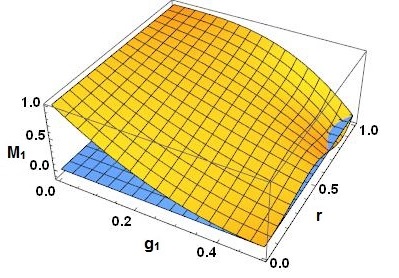}
\caption{Graph of parameters $g_1,r$ vs monogamy score $M_1$ for non generic GHZ class with $g_1\neq 0$,$g_2=0$,$g_3=0$,$r\in(0,1).$ }
\label{fig:2pvs3d}
\vspace{\floatsep}
\includegraphics[scale=0.5]{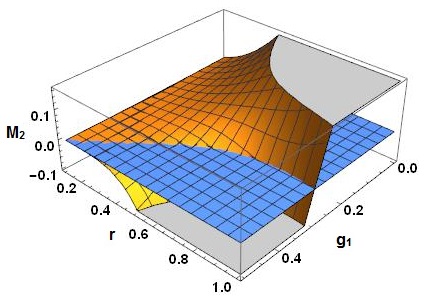}
\caption{Graph of parameters $g_1,r$ vs monogamy score $M_2$ for non generic GHZ class with $g_1\neq 0$,$g_2=0$,$g_3=0$,$r\in(0,1)$. }
\label{fig:2pva3d}
\end{figure}  

\hspace{1in}
$
 E_s= 
 \begin{cases} 
 1-2g_1(1-r) & \textrm{if  }r \in (0,1) \\  
 1-2g_1 & \textrm{if  } r = 1 
 \end{cases}
 $ 
 \hspace{0.5in}
$
E_a= 
 \begin{cases} 
 2(\frac{1}{2}-g_1)r & \textrm{if  }r \in (0,1) \\  
 1-2g_1 & \textrm{if  } r = 1 
 \end{cases}
 $\\
For $r\in(0,1)$, we have plotted the graphs of $M_1$ and $M_2$ in Fig. \ref{fig:2pvs3d} and Fig. \ref{fig:2pva3d} respectively. The blue plane corresponds to $M_1=0$ and $M_2=0$ for both Fig. \ref{fig:2pvs3d} and Fig. \ref{fig:2pva3d} respectively. Whereas the orange surface for Fig. \ref{fig:2pvs3d} corresponds to $E_s^2-E_{12}^2-E_{13}^2-E_{23}^2$ and for Fig. \ref{fig:2pva3d} corresponds to $E_{a}^2-E_{12}^2-E_{13}^2-E_{23}^2$. If the orange surface is above the blue plane then $M_1\geq0$ or $M_2\geq 0$ and monogamy relations $E_s^2\geq E_{12}^2+E_{13}^2+E_{23}^2$ or $E_a^2\geq E_{12}^2+E_{13}^2+E_{23}^2$ will hold. However, these two figures show that for specific regions of the parameters  $g_1$ and $r$, the values of $M_1$ and $M_2$ can be negative. We provide Fig. \ref{fig:2pvs} and Fig. \ref{fig:2pva}, in  appendix (\ref{appendix1}) that offer a better understanding of the monogamy relations \eqref{M1} and \eqref{M2}. The monogamy relations \eqref{M1} and \eqref{M2} are well satisfied(violated) for any pair $(g_1,r)$ in the gray(yellow) coloured region of Fig.  \ref{fig:2pvs} and Fig. \ref{fig:2pva} in appendix (\ref{appendix1}).
 
\begin{figure} [h]
\centering
\includegraphics[scale=.5]{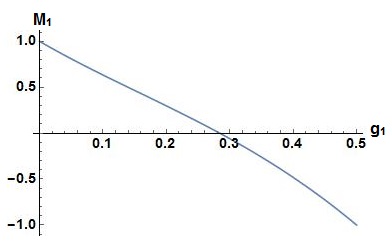}
\caption{Graph of parameter $g_1$ vs monogamy score $M_1(M_2)$ for non generic GHZ class with $g_1\neq 0$,$g_2=0$,$g_3=0$,$r=1$.  }
\label{fig:2pvr=1}
\end{figure}

Next, we explore monogamy when $r=1$.  We have plotted the graph of $M_1(\textrm{or } M_2)$ vs $g_1$ in Fig.  \ref{fig:2pvr=1}. From this figure, it is clear that inequalities \eqref{M1} and \eqref{M2} are satisfied for states corresponding to $0\leq g_1\leq 0.28\textrm{(approximately)}$ and violated for states corresponding to $0.28\leq g_1 <\frac{1}{2}$.

\textbf{Case 3 ($g_1\neq0$, $g_2\neq0$, $g_3=0$) :} In this case, the complexity of the non-generic GHZ class gradually increases. The concurrences of the reduced states are $C_{12}=0,C_{13}=\frac{2g_2\sqrt{1-4g_{1}^{2}}}{4k},C_{23}=\frac{2g_1\sqrt{1-4g_{2}^{2}}}{4k}.$ The source entanglement and accessible entanglement of $\ket{\psi(\vec{g},r)}$ are \\

\begin{figure} [h]
\centering
\includegraphics[scale=.5]{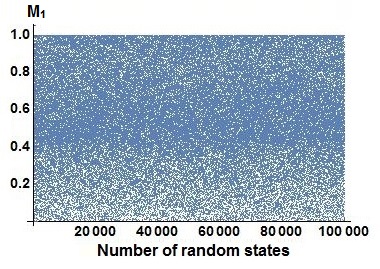}
\caption{Monogamy score $M_1$ for random states belonging to non generic GHZ class with $g_1\neq0$,$g_2\neq0$,$g_3=0$,$r\in(0,1)$. }
\label{fig:1pv}
\vspace{\floatsep}
\includegraphics[scale=.5]{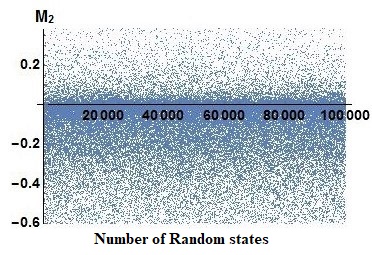}
\caption{Monogamy score $M_2$ for random states belonging to non generic GHZ class with $g_1\neq0$,$g_2\neq0$,$g_3=0$,$r\in(0,1)$.}
\label{fig:1pva}
\end{figure}

\hspace{.7in}
$
 E_s= 
 \begin{cases} 
 1-4g_1g_2(1-r) & \textrm{if  }r \in (0,1) \\  
 1-4g_1g_2 & \textrm{if  } r = 1 
 \end{cases}
 $
 \hspace{0.5in}
$
 E_a= 
 \begin{cases} 
 4(\frac{1}{2}-g_1)(\frac{1}{2}-g_2)r & \textrm{if  }r \in (0,1) \\   4(\frac{1}{2}-g_1)(\frac{1}{2}-g_2) & \textrm{if  } r = 1 
 \end{cases}
 $\\
Due to the challenges of providing an analytical proof when  $r\neq1$, we performed numerical simulations with $10^5$ random pure states. We have plotted the values of $M_1$ and $M_2$ for these random pure states in Fig.  \ref{fig:1pv} and Fig.  \ref{fig:1pva}, respectively. As $M_1\geq0$ in Fig.  \ref{fig:1pv} our numerical evidences suggest that $E_{s}^{2}\geq E_{12}^2+E_{13}^2+E_{23}^2$ in this case, whereas Fig.  \ref{fig:1pva} tells us that  the states from this case may violate $E_{a}^{2}\geq E_{12}^2+E_{13}^2+E_{23}^2$. In Appendix \ref{appendix2}, we consider the special case $g_1=g_2$ and plot the graph of $M_1$ vs $g_1,r$ in Figure \ref{fig:2app1pv=0}.

\begin{figure} [H]
\centering
\includegraphics[scale=.45]{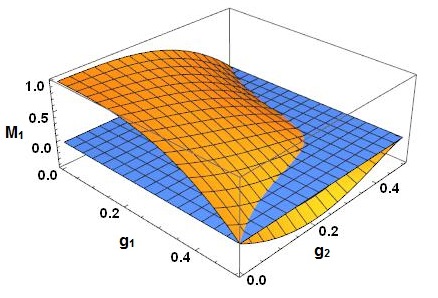}
\caption{Graph of parameters $g_1$,$g_2$ vs monogamy score $M_1$ for non generic GHZ class with $g_1\neq0$,$g_2\neq0$,$g_3=0$,$r=1$.}
\label{fig:3d1pvr=1}
\end{figure}

\begin{figure}
\centering
\includegraphics[scale=0.45]{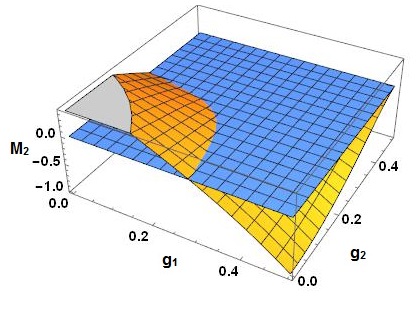}
\caption{Graph of parameters $g_1$,$g_2$ vs monogamy score $M_2$ for non generic GHZ class with $g_1\neq0$,$g_2\neq0$,$g_3=0$,$r=1$.}
\label{fig:3d1pvr=1a}
\end{figure}

For $r=1$, we have also plotted the graph of $M_1$ and $M_2$ in Fig.  \ref{fig:3d1pvr=1} and \ref{fig:3d1pvr=1a} respectively. These figures clearly show that monogamy relations \eqref{M1} and \eqref{M2} may be satisfied or violated depending upon the value of $(g_1,g_2)$. 

 For further clarity, figures  (\ref{fig:1pvr=1} and \ref{fig:1pvr=1a}), provided in Appendix \ref{appendix3} illustrate that states corresponding to ($g_1,g_2$) in the gray(yellow) region of these figures satisfy(violate) the inequalities \eqref{M1} and \eqref{M2} respectively.
\subsection{Generic GHZ states}
We now turn our attention to the generic states of the GHZ class. The source entanglement and accessible entanglement are given by:
\begin{equation}
    E_s=1-8g_1g_2g_3(1+f_z[\log(f_z)(1-\frac{1}{2}\log(f_z))-1]),
\end{equation}
\begin{equation}
    \label{genericaccessible}
    E_a=(\frac{1}{2}-g_1)(\frac{1}{2}-g_2)(\frac{1}{2}-g_3),
\end{equation}
where $g_i\in[0,\frac{1}{2})$ $\forall i=1(1)3$, $f_z=\frac{2|Re(z^2)|}{1+|z|^4}$ and $z\in\mathbb{C}$ with $|z|\leq1$ . The concurrences are given by the equations \eqref{c12}-\eqref{c23}.
\begin{figure} [H]
\centering
\includegraphics[scale=.5]{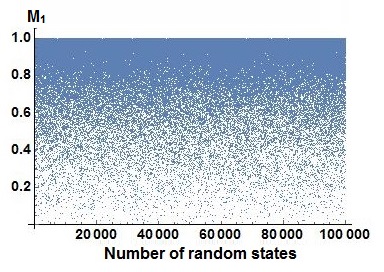}
\caption{Monogamy score $M_1$ for random states belonging to generic GHZ class.}
\label{fig:gens}
\end{figure}
\begin{figure}[h]
\centering
\includegraphics[scale=.5]{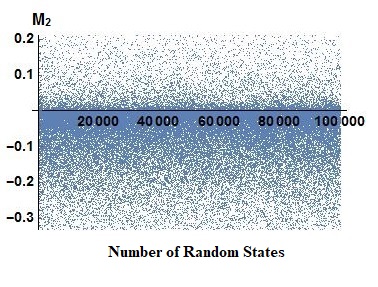}
\caption{Monogamy score $M_2$ for random states belonging to generic GHZ class.}
\label{fig:gena}
\end{figure} 
Due to the difficulties in the analytical proof of inequalities \eqref{M1} and  \eqref{M2}, we have done numerical simulation with $10^5$ random pure states from the generic class for testing the validity of eqn. \eqref{M1} and \eqref{M2}. The values of $M_1$ and $M_2$ for these random states are plotted in Fig.  \ref{fig:gens} and Fig. \ref{fig:gena}. In Fig.  \ref{fig:gens}, we can see that $M_1\geq0$ always. So, our numerical study suggests that generic GHZ states satisfy the monogamy relation $E_{s}^{2}\geq E_{12}^{2}+E_{13}^{2}+E_{23}^{2}$. But the inequality $E_a^2\geq E_{12}^2+E_{13}^2+E_{23}^2$ can be violated sometimes by generic GHZ states as evident from the Fig. \ref{fig:gena}. Further analysis on some particular cases of generic states of GHZ class has been done in Appendix \ref{appendix4} to check the monogamy inequality \eqref{M1}. 

\subsection{\texorpdfstring{States from $MES_3$ of the form $\frac{1}{\sqrt{k}}g_{x}^{1}\otimes g_{x}^{2}\otimes g_{x}^{3}\ket{GHZ}$ where  $g_1,g_2,g_3$ are non-zero}{} }
\begin{figure} [H]
\centering
\includegraphics[scale=.5]{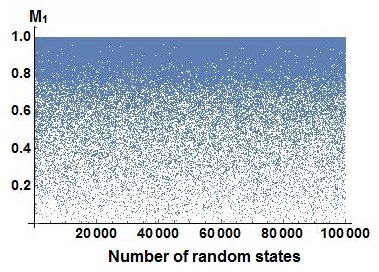}
\caption{Monogamy score $M_1$ for random states belonging to $\frac{1}{\sqrt{k}}g_{x}^{1}\otimes g_{x}^{2}\otimes g_{x}^{3}\ket{GHZ}$.}
\label{fig:mes}
\includegraphics[scale=.5]{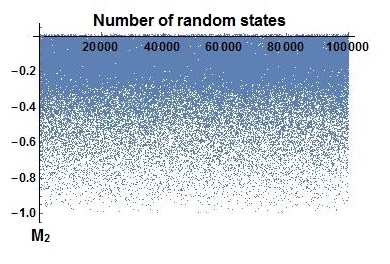}
\caption{Monogamy score $M_2$ for random states belonging  to $\frac{1}{\sqrt{k}}g_{x}^{1}\otimes g_{x}^{2}\otimes g_{x}^{3}\ket{GHZ}$.}
\label{fig:ames}
\end{figure}

Next, we  we examine the states $\frac{1}{\sqrt{k}}g_{x}^{1}\otimes g_{x}^{2}\otimes g_{x}^{3}\ket{GHZ}$, $\vec{g}\neq \vec{0}$, that belong to the $MES_3$.  For these states, the source entanglement is $E_s=1$,  while the accessible entanglement remains the same as given by equation eqn. \eqref{genericaccessible}. The concurrences of three reduced subsystems can be easily obtained from relations \eqref{c12}-\eqref{c23} by substituting $|z|=1$, allowing us to calculate the corresponding entanglement of formation using equation \eqref{Ef}. Numerical simulations using $10^5$ pure states from this class have been conducted, as shown in Fig. \ref{fig:mes} and Fig. \ref{fig:ames}. From Fig. \ref{fig:mes}, we observe that $E_{s}^{2}\geq E_{12}^{2}+E_{13}^{2}+E_{23}^{2}$  (see Appendix \ref{appendix5} for a particular case $g_1=g_2=g_3\neq 0$). On the other hand, figure \ref{fig:ames} reveals that the relation   $E_{a}^{2}\geq E_{12}^{2}+E_{13}^{2}+E_{23}^{2}$ is violated by almost all states considered from this class.
We will now extend our investigation to the W class of state to complete our study of three-qubit genuinely entangled pure states.

\section{Multipartite monogamy in  W class} 
\begin{figure} [H]
\centering
\includegraphics[scale=.5]{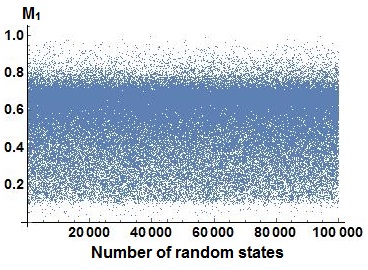}
\caption{Monogamy score $M_1$ for random states belonging to W class.}
\label{fig:ws}
\end{figure}

\begin{figure}[h]
\centering
\includegraphics[scale=.5]{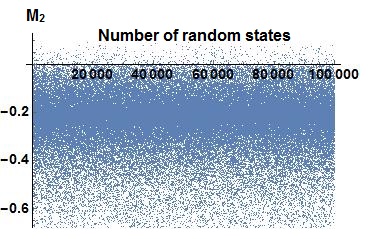}
\caption{Monogamy score $M_2$ for random states belonging to W class.}
\label{fig:wa}
\end{figure}

Any state in the W class (up to local unitary) can be represented as $$\ket{\psi}=\sqrt{t}\ket{000}+\sqrt{x}\ket{100}+\sqrt{y}\ket{010}+\sqrt{z}\ket{001}$$ where $x,y,z>0$, $t\geq0$ and $x+y+z+t=1$. A state in W class can belongs to $MES_3$ if and only if $t=0$. The concurrences of the reduced states are $C_{12}=2\sqrt{xy}$, $C_{13}=2\sqrt{xz}$, $C_{23}=2\sqrt{yz}$. All states in the W class have zero tangles. Now the source entanglement is $E_s=1-t^3$, and accessible entanglement is $E_a=27xyz$. The numerical simulations with $10^5$ random pure states from this class have been executed, and Fig. \ref{fig:ws} clearly suggests that $E_s^2\geq E_{12}^2+E_{13}^2+E_{23}^2$ always hold. When we use accessible entanglement as the upper bound, a drastic change in monogamy nature has been found. Numerical simulations (Fig. \ref{fig:wa}) show that in almost all of the cases, the monogamy relation \eqref{M2} is violated.

Finally we have considered  $\ket{W}$ state which is $\ket{W}=\frac{1}{\sqrt{3}}(\ket{100}+\ket{010}+\ket{001})$ and $\ket{W}\in \text{MES}_3$. Here, we have $E_{12}=E_{13}=E_{23}\approx 0.550048$ and $E_s=E_a=1$. Therefore $M_1=M_2\approx 0.0923424> 0$.   

\section{Observations and Interpretations}
The presence of monogamy restricts the free sharing of entanglement. We provide a detailed tabular overview of the distribution of bipartite entanglement in the reduced systems for GHZ and W classes. Our results indicate that inequality \eqref{M1} is satisfied in two scenarios within the GHZ class: when none of the three reduced subsystems are entangled and when all three subsystems have non-zero entanglement. Notably, violation of inequality  \eqref{M1} only occurs when one or two reduced subsystems lack entanglement. Our study clearly shows that such violations are influenced by the parameters $g_i \textrm{ and } r$, which are directly related to the entanglement of the reduced subsystems, as seen from Equations \eqref{c12}-\eqref{c23}. Specifically, when one or two $g_i$'are zero, monogamy \eqref{M1} is violated in certain regions. The parameter $r$ also plays a crucial role in these violations. The monogamy relation exhibits more consistent violations for $r=1$ instead of $0<r<1$. Notably, in cases where only one  $g_i$ is zero, no violation occurs due to the term  $\frac{2|r|^2}{1+|r|^4}$ in the expression for concurrences and the entanglement of formation of the reduced subsystems. This term reaches its maximum value $1$  when $r=1$, suggesting that violations are less likely for $r\neq 1$.  For the W class, where all three reduced subsystems exhibit non-zero entanglement always, Fig. \ref{fig:ws} indicates that monogamy \eqref{M1} are generally satisfied. Regarding inequality \eqref{M2}, it is satisfied only when there is no bipartite entanglement within any of the reduced subsystems. Violations of relation \eqref{M2} occur when at least one subsystem exhibits bipartite entanglement, with more violations arising as more reduced subsystems display bipartite entanglement. The status of monogamy relations for three-qubit states is summarized in Tables \ref{table1} and \ref{table2}. 

\begin{widetext}
\begin{table*}[ht]
\begin{tabular}{|c|c|c|c|c|}
\hline
\multicolumn{3}{|c|}{\begin{tabular}[c]{@{}c@{}}\textbf{Three qubit Genuine}\\ \textbf{Multipartite entangled states}\end{tabular}}                                                                                               & \begin{tabular}[c]{@{}c@{}}\textbf{Status of multipartite entanglement} \\   \textbf{in reduced subsystems 12, 13, 23}\end{tabular}                                                        & \multicolumn{1}{c|}{$E_{s}^{2}\geq E_{12}^{2}+E_{13}^{2}+E_{23}^{2}$}                                                         \\ \hline
\multirow{6}{*}{\begin{tabular}[c]{@{}c@{}}Non-generic \\ GHZ class states\end{tabular}} & \multirow{2}{*}{\begin{tabular}[c]{@{}c@{}}$g_1=0,g_2=0,g_3=0$\end{tabular}}         & $r\in(0,1)$                      & \multirow{2}{*}{\begin{tabular}[c]{@{}c@{}}No bipartite entanglement exists\\ between reduced subsystems 12, 13, 23 \end{tabular}}                                          & Well satisfied                                                                                            \\ \cline{3-3} \cline{5-5} 
                                                                               &                                                                                              & $r=1$                            &                                                                                                                                                                        & Well satisfied                                                                                            \\ \cline{2-5} 
                                                                               & \multirow{2}{*}{\begin{tabular}[c]{@{}c@{}}$g_1\neq 0, g_2=0,g_3=0$\end{tabular}}     & \multicolumn{1}{l|}{$r\in(0,1)$} & \multirow{2}{*}{\begin{tabular}[c]{@{}c@{}}Bipartite entanglement exists between \\ 23 but no bipartite entanglement exists\\  between reduced subsystems 12,13\end{tabular}} & \begin{tabular}[c]{@{}c@{}}Partially satisfied in the gray \\ region of $g_1$ and $r$ in figure \ref{fig:2pvs}\end{tabular}   \\ \cline{3-3} \cline{5-5} 
                                                                               &                                                                                              & $r=1$                            &                                                                                                                                                                        & \begin{tabular}[c]{@{}c@{}}Partially satisfied when \\ $g_1<0.28$ in figure \ref{fig:2pvr=1}\end{tabular}           \\ \cline{2-5} 
                                                                               & \multirow{2}{*}{\begin{tabular}[c]{@{}c@{}}$g_1\neq 0,g_2\neq 0,g_3=0$\end{tabular}}   & \multicolumn{1}{l|}{$r\in(0,1)$} & \multirow{2}{*}{\begin{tabular}[c]{@{}c@{}}Bipartite entanglement exists between \\ 13, 23 but no bipartite entanglement\\exists between reduced subsystem 12 \end{tabular}} & Well satisfied (see figure \ref{fig:1pv})                                                                                           \\ \cline{3-3} \cline{5-5} 
                                                                               &                                                                                              & $r=1$                            &                                                                                                                                                                        & \begin{tabular}[c]{@{}c@{}}Partially satisfied in the gray \\ region of $g_1$ and $g_2$ in figure \ref{fig:1pvr=1}\end{tabular} \\ \hline
\multicolumn{3}{|c|}{Generic GHZ class states}                                                                                                                                                                              & \multicolumn{1}{c|}{\begin{tabular}[c]{@{}c@{}}Bipartite entanglement exists in all\\ reduced  subsystems 12, 13, 23\end{tabular}}                                              & Well satisfied (see figure \ref{fig:gens})                                                                                           \\ \hline
\multicolumn{3}{|c|}{\begin{tabular}[c]{@{}c@{}}State from GHZ class of the form \\ $\frac{1}{\sqrt{k}}g_{x}^{1}\otimes g_{x}^{2}\otimes g_{x}^{3}\ket{GHZ}$ where $g_1,g_2,g_3 \neq 0$\end{tabular}} & \begin{tabular}[c]{@{}c@{}}Bipartite entanglement exists in \\ all reduced subsystems 12, 13, 23\end{tabular}                                                            & Well satisfied (see figure \ref{fig:mes})                                                                                           \\ \hline
\multicolumn{3}{|c|}{W class states}                                                                                                                                                                             & \begin{tabular}[c]{@{}c@{}}Bipartite entanglement exists in \\ all reduced subsystems 12, 13, 23\end{tabular}                                                            & Well satisfied (see figure \ref{fig:ws})                                                                                           \\ \hline
\end{tabular}
\caption{Detailed analysis of proposed monogamy using source entanglement as multiparty entanglement measure in three-qubit genuine multipartite entangled pure states.}
\label{table1}
\end{table*}

\begin{table}[h]
\begin{tabular}{|c|c|c|c|c|}
\hline
\multicolumn{3}{|c|}{\begin{tabular}[c]{@{}c@{}}\textbf{Three qubit Genuine}\\ \textbf{Multipartite entangled states}\end{tabular}}                                                                                               & \begin{tabular}[c]{@{}c@{}}\textbf{Status of multipartite entanglement} \\   \textbf{in reduced subsystems 12, 13, 23}\end{tabular}                                                        & \multicolumn{1}{c|}{$E_{a}^{2}\geq E_{12}^{2}+E_{13}^{2}+E_{23}^{2}$}                                                         \\ \hline
\multirow{6}{*}{\begin{tabular}[c]{@{}c@{}}Non-generic \\ GHZ class states\end{tabular}} & \multirow{2}{*}{\begin{tabular}[c]{@{}c@{}}$g_1=0,g_2=0,g_3=0$\end{tabular}}         & $r\in(0,1)$                      & \multirow{2}{*}{\begin{tabular}[c]{@{}c@{}}No bipartite entanglement exists\\ between reduced subsystems 12, 13, 23 \end{tabular}}                                          & Well satisfied                                                                                            \\ \cline{3-3} \cline{5-5} 
                                                                               &                                                                                              & $r=1$                            &                                                                                                                                                                        & Well satisfied                                                                                            \\ \cline{2-5} 
                                                                               & \multirow{2}{*}{\begin{tabular}[c]{@{}c@{}}$g_1\neq 0, g_2=0,g_3=0$\end{tabular}}     & \multicolumn{1}{l|}{$r\in(0,1)$} & \multirow{2}{*}{\begin{tabular}[c]{@{}c@{}}Bipartite entanglement exists between \\ 23 but no bipartite entanglement exists\\  between reduced subsystems 12,13\end{tabular}} & \begin{tabular}[c]{@{}c@{}}Partially satisfied in the gray \\ region of $g_1$ and $r$ in figure \ref{fig:2pva}\end{tabular}   \\ \cline{3-3} \cline{5-5} 
                                                                               &                                                                                              & $r=1$                            &                                                                                                                                                                        & \begin{tabular}[c]{@{}c@{}}Partially satisfied when \\ $g_1<0.28$ in figure \ref{fig:2pvr=1}\end{tabular}           \\ \cline{2-5} 
                                                                               & \multirow{2}{*}{\begin{tabular}[c]{@{}c@{}}$g_1\neq 0,g_2\neq 0,g_3=0$\end{tabular}}   & \multicolumn{1}{l|}{$r\in(0,1)$} & \multirow{2}{*}{\begin{tabular}[c]{@{}c@{}}Bipartite entanglement exists between \\ 13, 23 but no bipartite entanglement\\exists between reduced subsystem 12 \end{tabular}} & partially satisfied (see figure \ref{fig:1pva})                                                                                            \\ \cline{3-3} \cline{5-5} 
                                                                               &                                                                                              & $r=1$                            &                                                                                                                                                                        & \begin{tabular}[c]{@{}c@{}}Partially satisfied in the gray \\ region of $g_1$ and $g_2$ in figure \ref{fig:1pvr=1a}\end{tabular} \\ \hline
\multicolumn{3}{|c|}{Generic GHZ class states}                                                                                                                                                                              & \multicolumn{1}{c|}{\begin{tabular}[c]{@{}c@{}}Bipartite entanglement exists in all\\ reduced  subsystems 12, 13, 23\end{tabular}}                                              & Almost all states violate 
(see figure \ref{fig:gena})                                                                                            \\ \hline
\multicolumn{3}{|c|}{\begin{tabular}[c]{@{}c@{}}State from GHZ class of the form \\ $\frac{1}{\sqrt{k}}g_{x}^{1}\otimes g_{x}^{2}\otimes g_{x}^{3}\ket{GHZ}$ where $g_1,g_2,g_3 \neq 0$\end{tabular}} & \begin{tabular}[c]{@{}c@{}}Bipartite entanglement exists in \\ all reduced subsystems 12, 13, 23\end{tabular}                                                            & Almost all states violate (see figure \ref{fig:ames})                                                                                             \\ \hline
\multicolumn{3}{|c|}{W class states}                                                                                                                                                                             & \begin{tabular}[c]{@{}c@{}}Bipartite entanglement exists in \\ all reduced subsystems 12, 13, 23\end{tabular}                                                            & Almost all states violate (see figure \ref{fig:wa})                                                                                          \\ \hline
\end{tabular}
\caption{Detailed analysis of proposed monogamy using accessible entanglement as multiparty entanglement measure in three-qubit genuine multipartite entangled pure states }
\label{table2}
\end{table}
\end{widetext}

\section{Conclusion}
In this work, our primary goal is to explore the distribution of entanglement through the lens of multipartite monogamy relations. We investigated whether squared source entanglement or squared accessible entanglement can serve as an upper bound for the sum of the squares of the entanglement of formation in all possible reduced bipartite states within pure three-qubit genuinely entangled states. Except for a few cases in non-generic GHZ class states, the source entanglement ($E_s$) monogamy relation,  $E_{s}^{2}\geq E_{12}^2+E_{13}^2+E_{23}^2$  generally satisfied for pure GHZ and W class states. We found that violations in non-generic GHZ pure states occur due to the lack of entanglement in one or two reduced subsystems, while bipartite entanglement is concentrated in the remaining subsystems. This leads to scenarios where the sum of the squared entanglement of formation can exceed the squared source entanglement. Conversely, in genuinely multipartite entangled three-qubit pure states, where each reduced subsystem has either non-zero or zero entanglement, the monogamy $E_{s}^{2}\geq E_{12}^2+E_{13}^2+E_{23}^2$ holds true. However, the monogamy relation $E_{a}^{2}\geq E_{12}^2+E_{13}^2+E_{23}^2$ behaves differently from $E_{s}^{2}\geq E_{12}^2+E_{13}^2+E_{23}^2$. The monogamy inequality involving accessible entanglement ($E_a$) is satisfied only when there is no bipartite entanglement in any of the reduced subsystems; otherwise, it is always violated. These observations align with familiar physical concepts. The violation of inequality $E_{s}^{2}\geq E_{12}^2+E_{13}^2+E_{23}^2$ may signal the absence of entanglement in one or two reduced subsystems. This concept could be extended as a new research direction for higher-dimensional qubit systems and mixed states. We hope our study will contribute to further research in multipartite entanglement distribution and potentially enhance applications in quantum key distribution and quantum cryptography.\\

\section*{ACKNOWLEDGEMENTS}
Priyabrata Char acknowledges the support from the Department of Science and Technology(Inspire), New Delhi, India. The authors D. Sarkar and I. Chattopadhyay acknowledge it as a Quest initiative.

\appendix
\section{\texorpdfstring{Non-Generic GHZ state with with $g_1\neq 0,g_2=0,g_3=0,r\in(0,1)$}{}}
\label{appendix1}
\begin{figure} [h]
\centering
\includegraphics[scale=.4]{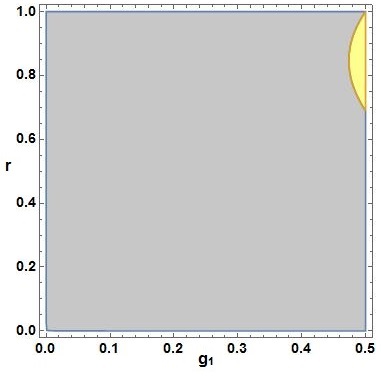}
\caption{Non-Generic GHZ state with $g_1\neq 0, g_2=0, g_3=0,r\in(0,1)$.  Monogamy relations w.r.t. source entanglement are held in the gray region and are violated in the yellow region.}
\label{fig:2pvs}
\vspace{\floatsep}
\includegraphics[scale=.4]{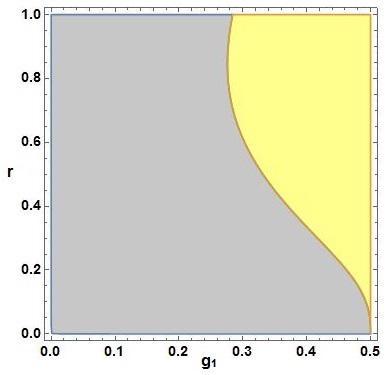}
\caption{Non-Generic GHZ state with $g_1\neq 0, g_2=0, g_3=0,r\in(0,1)$.  Monogamy relations w.r.t. accessible entanglement are held in the gray region and are violated in the yellow region.}
\label{fig:2pva}
\end{figure}

Monogamy relations of source and accessible entanglement for non-generic GHZ state are shown through Fig. \ref{fig:2pvs} and \ref{fig:2pva} respectively. The gray region in Fig. \ref{fig:2pvs} and Fig. \ref{fig:2pva} represents the pair $(g_1,r)$ where the orange surface is above the blue plane in Fig. \ref{fig:2pvs3d} and \ref{fig:2pva3d} respectively. On the other hand, the yellow region in Fig. \ref{fig:2pvs} and Fig. \ref{fig:2pva} represents the pair $(g_1,r)$ where the orange surface is below the blue plane in Fig. \ref{fig:2pvs3d} and \ref{fig:2pva3d} respectively. 
\section{\texorpdfstring{Non-Generic GHZ class with $g_1\neq 0,g_2\neq 0, g_3=0,r\in(0,1)$.}{}}
\label{appendix2}
\begin{figure}[H] 
\centering
\includegraphics[scale=.5]{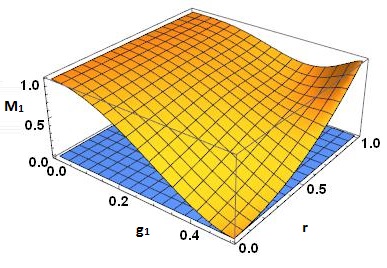}
\caption{Graph of parameters $g_1,r$ vs monogamy score $M_1$ for non generic GHZ class states with $g_1=g_2\neq 0,g_3=0,r\in(0,1).$ }
\label{fig:2app1pv=0}
\end{figure}

We are now focusing on the non-generic GHZ class where one parameter, $g_3=0$ is set to zero, and specifically considering the case where  $g_1=g_2\in(0,\frac{1}{2}]$. We calculate $M_1$ and present the results in Fig. \ref{fig:2app1pv=0}. In this figure, the blue plane represents $M_1=0$ while the orange surface, which is consistently above the blue plane, corresponds to $M_1=E_{s}^{2}-E_{12}^{2}-E_{13}^{2}-E_{23}^{2}$. This clearly demonstrates the validity of monogamy for source entanglement.

\section{\texorpdfstring{Non-Generic GHZ class with $g_1\neq 0,g_2\neq 0,g_3=0,r=1$}{}}
\label{appendix3}
\begin{figure} [H]
\centering
\includegraphics[scale=.44]{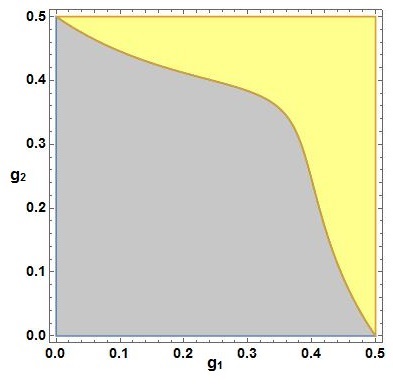}
\caption{Non generic GHZ class with $g_1\neq 0, g_2\neq0, g_3=0, r=1$. Monogamy relation w.r.t source entanglement holds In the gray region and is violated in the yellow region.}
\label{fig:1pvr=1}
\end{figure}

\begin{figure}[H]
\centering
\includegraphics[scale=0.44]{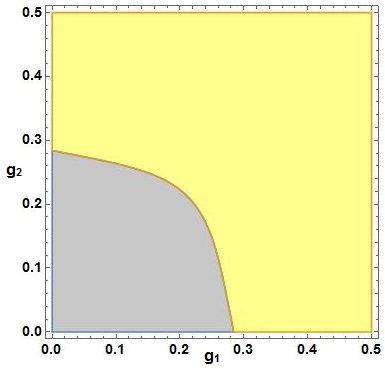}
\caption{Non generic GHZ class $g_1\neq 0, g_2\neq0, g_3=0,r=1$. Monogamy relation w.r.t accessible entanglement holds In the gray region and is violated in the yellow region.}
\label{fig:1pvr=1a}
\end{figure}

Status of two monogamy relations \eqref{M1} and \eqref{M2} for the Non-Generic GHZ class with $g_1\neq 0,g_2\neq 0,g_3=0,r=1$ corresponding to source and accessible entanglement are shown in Fig \ref{fig:1pvr=1} and \ref{fig:1pvr=1a}.
Gray region in Fig. \ref{fig:1pvr=1} and Fig. \ref{fig:1pvr=1a} represents the pair $(g_1,g_2)$ for which the orange surface is above the blue plane in Fig. \ref{fig:3d1pvr=1} and \ref{fig:3d1pvr=1a} respectively. On the other hand yellow region in Fig. \ref{fig:1pvr=1} and Fig. \ref{fig:1pvr=1a} represents the pair $(g_1,g_2)$ for which the orange surface is below the blue plane in Fig. \ref{fig:3d1pvr=1} and \ref{fig:3d1pvr=1a}.
\section{Generic GHZ class}
\label{appendix4}
\begin{figure}[h]
\centering
\includegraphics[scale=.45]{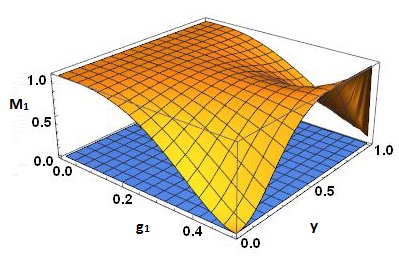}
\caption{Graph of $g_1,y$ vs $M_1$ for non generic GHZ class states with $g_1=g_2=g_3$ and $z=iy.$}
\label{fig:3appcs1}
\end{figure}

\begin{figure}[H]
\centering
\includegraphics[scale=.45]{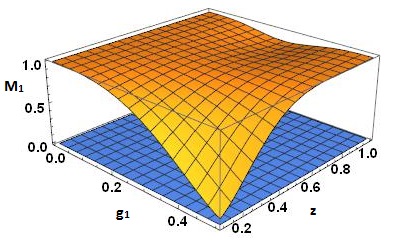}
\caption{Graph of $M_1$ for non generic GHZ class states with $g_1=g_2=g_3$ and $z$ is real.}
\label{fig:3appcs2}
\end{figure}

\begin{figure} [H]
\centering
\includegraphics[scale=.45]{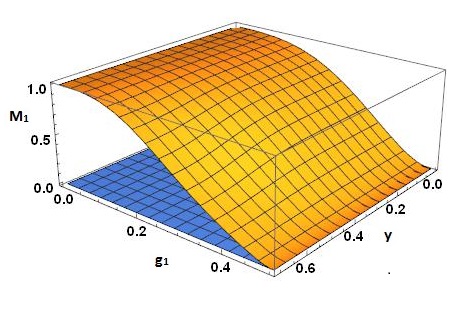}
\caption{Graph of $M_1$ for non generic GHZ class states with $g_1=g_2=g_3$ and $z=(\pm1+i)y$.}
\label{fig:3appcs3}
\end{figure}

Here, we will examine monogamy \eqref{M1} for three specific cases of generic states in the GHZ class. For each case, we calculate $M_1=E_s^2-E_{12}^{2}-E_{13}^{2}-E_{23}^{2}$ and plot them in Fig. \ref{fig:3appcs1}, \ref{fig:3appcs2}, \ref{fig:3appcs3}. The blue plane corresponds to $M_1=0$ and the orange surface corresponds to $M_1=E_{s}^{2}-E_{12}^{2}-E_{13}^{2}-E_{23}^{2}$. If the orange surface is above the blue plane then $M_1\geq0$ and monogamy relation $E_s^2\geq E_{12}^2+E_{13}^2+E_{23}^2$ will hold.\\
 
\textsl{Case 1}: We consider $g_1=g_2=g_3\in(0,\frac{1}{2}]$ and $z=iy$ is a purely imaginary number where $|z|\leq 1$. We plot $M_1$ vs $g_1,y$ in Fig.  \ref{fig:3appcs1}, where the orange surface is always above the blue plane. Hence $M_1\geq0$, which in turns implies $E_s^2\geq E_{12}^2+E_{13}^2+E_{23}^2$.

\textsl{Case 2}: Let $g_1=g_2=g_3\in(0,\frac{1}{2}]$ and $z$ is a real number with $|z|\leq 1$.  We plot $M_1$ vs $g_1,z$ in Fig. \ref{fig:3appcs2} where the orange surface is always above the blue plane. Hence $M_1\geq0$, this implies that  $E_s^2\geq E_{12}^2+E_{13}^2+E_{23}^2$.

\textsl{Case 3}:
Let $g_1=g_2=g_3\in(0,\frac{1}{2}]$ and $Re(z^2)=0$. Then $z=(\pm1+i)y$ where $y$ is a positive real number with $y\leq\frac{1}{\sqrt{2}}$ so that $|z|\leq1$. then $f_z=0$ and $E_s=1-8g_{1}^{3}$. We plot $M_1$ vs $g_1,y$ graph in Fig. \ref{fig:3appcs3} where the orange surface is always above the blue plane. Hence $M_1\geq0$, this implies that  $E_s^2\geq E_{12}^2+E_{13}^2+E_{23}^2$.
 
\section{\texorpdfstring{States from $MES_3$ of the form $\frac{1}{\sqrt{k}}g_{x}^{1}\otimes g_{x}^{2}\otimes g_{x}^{3}\ket{GHZ}$ where  $g_1,g_2,g_3\neq 0$}{}}
\label{appendix5}
\begin{figure} [H]
\centering
\includegraphics[scale=.5]{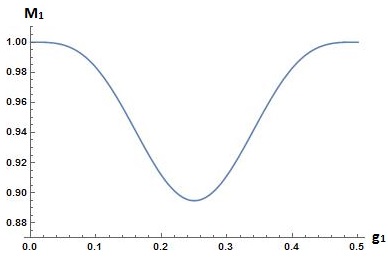}
\caption{Graph of $g_1$ vs $M_1$ for the states of the form $\frac{1}{\sqrt{k}}g_{x}^{1}\otimes g_{x}^{2}\otimes g_{x}^{3}\ket{GHZ}$ with $g_1=g_2=g_3.$}
\label{fig:4app}
\end{figure}

We consider a special case for the states from $\text{MES}_3$ which are of the form $\frac{1}{\sqrt{k}}g_{x}^{1}\otimes g_{x}^{2}\otimes g_{x}^{3}\ket{GHZ}$ where  $g_1=g_2=g_3\neq 0$. Then We have plotted the graph of $M_1$ vs $g_1$ in Fig. \ref{fig:4app} which shows that $M_1\geq0$ and hence $E_s^2\geq E_{12}^2+E_{13}^2+E_{23}^2$. 
\end{document}